# THE MODEL OF DRYING SESSILE DROP OF COLLOIDAL SOLUTION

I.V.Vodolazskaya, Yu.Yu.Tarasevich

*Astrakhan State University, 20a Tatishchev St.,*

*Astrakhan, 414056, Russia*

e-mail: tarasevich@aspu.ru

We have proposed and investigated a model of drying colloidal suspension drop placed onto a horizontal substrate in which the sol to gel phase transition occurs. The temporal evolution of volume fraction of the solute and the gel phase dynamics were obtained from numerical simulations. Our model takes into account the fact that some physical quantities are dependent on volume fraction of the colloidal particles.

## 1. Introduction

Drying sessile drops have been studied experimentally by a number of groups. The contact line pinning, the redistribution of the solute particles, ring formation, sol–gel transition, the drop surface deformation were investigated in the works.[1–4] Understanding of the process of drying of solutions is important for scientific, industrial and medical applications.

Several models[1, 5–10] have been proposed for the deposit formation and the motion of the liquid-deposit boundary inside a drop. Authors investigated one-dimensional problem utilizing the matter conservation and some geometric assumptions about the drop–atmosphere interface or the velocity of the phase boundary. In the work[1] authors assumed that contact line pinning and evaporation are sufficient conditions for the deposit ring formation. When evaporation removes liquid from the contact line, a flow arises to keep the substrate wet up to that point. The solute in the drop is dragged to the contact line by this flow, where it accumulates. The solute particles stop moving when its local volume fraction $W$ reaches the threshold value $W_f$. A zone of immobilized solute (deposit phase) appears at the perimeter and begins to move inward. In the models[5,6,9,10] the deposit phase does not alter the outward liquid flows and evaporation flow. In the works[7,8] authors suggested that fixed solute stops outward flows and evaporation flux from the surface of the deposit region. In our opinion, models proposed in the works[5,6,9,10] are more adapted to inorganic deposition and models[7, 8] are adapted to colloidal gels.

Now the most complicated problem relates to determining the rate of evaporation from the free surface of the drop. Well-known functional forms of the evaporation rate $J$ [1, 11, 12] do not take into account the complex geometry of drop surface and the drop composition. Thus, many researchers [6, 8] use in their calculations the uniform evaporation rate $J = \text{const}$ or evaporation rate modifications.[7]

## 2. Model and assumptions

Our model is applicable to drops of aqueous protein solution. A drop is deposited onto a horizontal substrate under usual and uniform environment and is single-phase (sol phase), but some time after the evaporation begins a second phase appears in the system (gel phase). We assume the volume fraction $W_f$ of the solute in the gel phase is high and its value is fixed. In the sol phase, space averaged volume fraction $W$ of the solute changes with time.

An axisymmetric thin drop has radius in the substrate plane, $R$, the height of the drop is $h(r,t)$ ($r$ is the radial coordinate or distance from the drop center, $t$ is time), and we assume that $h(r,t) \ll R$.

During the evaporation process, the contact line does not recede; which means the quantity $R$ is constant.

We assume the drop is rather small, hence surface tension is dominant, and the gravitational effects can be neglected (the ratio $\rho g R^2 / 2\sigma$ is smaller than 0.25;[5] here $g$ is the gravitational constant, $\sigma$ is the surface tension at the liquid–air interface).

If the colloidal particles are large enough, diffusion can be assumed negligible. We assume that the fluid density $\rho$ is constant and equals the solute density. Hence the sedimentation is negligible. In such a way the particle velocity equals the fluid velocity $\vec{\upsilon}$. It is known that the diffusion coefficient of albumin molecules in water is very small: $D_a = 5.5 \cdot 10^{-11} m^2/s$,[13] for the comparison, the ones for the salt NaCl is $D_s = 1.2 \cdot 10^{-9} m^2/s$.[14]

We can utilize the lubrication approximation[15] for the thin drops and slow flows. Fluid flow is governed by the Navier–Stokes equations:

$$0 = -\frac{\partial p}{\partial r} + \eta \frac{\partial^2 u}{\partial z^2},$$
$$0 = -\frac{\partial p}{\partial z} \tag{1}$$

with the boundary conditions (2) for the pressure $p$ and the radial fluid velocity $u$ ($z$ is the vertical coordinate or distance under the substrate, $\eta$ is the viscosity):

$$p\big|_{z=h} = -\frac{\sigma}{r}\frac{\partial\left(r\frac{\partial h}{\partial r}\right)}{\partial r}, \quad \frac{\partial u}{\partial z}\bigg|_{z=h} = 0, \quad u\big|_{z=0} = 0. \tag{2}$$

If set of equations (1) are integrated over the boundary conditions (2), the radial velocity and then the height–averaged velocity can be written:

$$u = -\frac{\sigma}{\eta}\frac{\partial}{\partial r}\left(\frac{1}{r}\frac{\partial}{\partial r}\left(r\frac{\partial h}{\partial r}\right)\right)\left(\frac{z^2}{2} - hz\right),$$

$$<u> = \frac{1}{h}\int_0^h u\,dz = \frac{h^2}{3}\frac{\sigma}{\eta}\frac{\partial}{\partial r}\left(\frac{1}{r}\frac{\partial}{\partial r}\left(r\frac{\partial h}{\partial r}\right)\right). \quad (3)$$

The fluid and the solute conservations (4) in the drop with the boundary and initial conditions (5) give a possibility to obtain the height $h(r,t)$ and height averaged volume fraction $W(r,t)$ of the solute:

$$\frac{\partial h}{\partial t} = -\frac{1}{r}\frac{\partial(rh\langle u\rangle)}{\partial r} - \frac{J}{\rho},$$

$$\frac{\partial(hW)}{\partial t} = -\frac{1}{r}\frac{\partial}{\partial r}\left(rWh\langle u\rangle\right). \quad (4)$$

$$h\big|_{t=0} = h_0\left(1 - \left(\frac{r}{R}\right)^2\right),$$

$$\frac{\partial h}{\partial r}\bigg|_{r=0} = 0, \quad h\big|_{r=R} = 0,$$

$$\langle u\rangle\big|_{r=0} = \langle u\rangle\big|_{r=R} = 0, \quad (5)$$

$$W\big|_{t=0} = W_0, \quad \frac{\partial W}{\partial r}\bigg|_{r=0} = 0.$$

Here $J$ is the functional form of the evaporation rate, $W_0$ is the initial volume fraction of the solute (the solute was assumed to be uniformly distributed throughout the drop at $t = 0$), $h_0 = h(0,0)$ and initial drop shape is equilibrium.

## 2.1. *The sol–gel transition*

It is known that the gelation process of the protein molecules accelerates when the solute volume fraction $W$ increases. The gel phase occurs after the solute volume fraction $W$ has reached the specific value $W_f$. The admixtures in the solution affect the gelation process. The increase of salt content in the solution screens out the electrostatic interaction between the molecules, and the solution gelation accelerates.[16] The salt quantity in the solution is described as ionic strength value. We came to a conclusion that phase transition sol–gel is the gradual concentration transition. Since gel is a solid state,[6] the protein molecules stop moving and gel region keeps its shape. The fluid molecules place among protein molecules or its aggregates and create the connections. Hence it is necessary to include in the model the appearance of the new phase with different physical properties. The presence of the gel phase must affect both the hydrodynamics and the evaporation. In our model the gelation process in sol zone near the liquid–air interface is not taken into account. It is known, that the sol–gel transition is accompanied with the divergence of the viscosity.[4] We suggest for the viscosity $\eta$ the functional form with the divergence if $W \approx W_f$, for example:

$$f = \frac{\eta}{\eta_0} = \exp\left(a\left(\frac{W}{W_f}\right)^b\right). \tag{6}$$

Figure 1 shows $f$ as a function of $W/W_f$. Here $\eta_0$ is the viscosity of the pure solvent and $a$ and $b$ are natural numbers. It is easy to see that the value of the parameters $a$ and $b$ are dependent of the protein type and the ionic strength value.

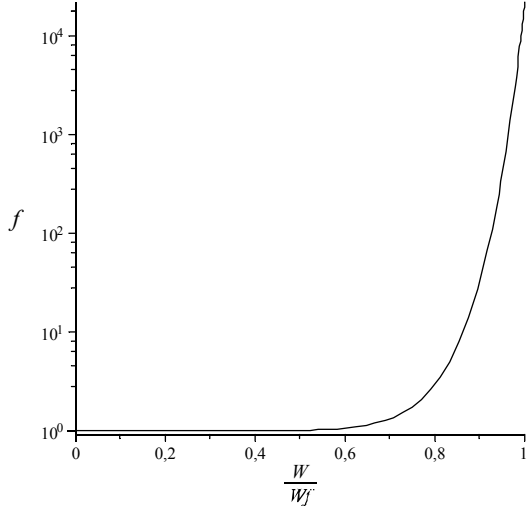

Figure 1. The quantity $f = \eta/\eta_0$ as a function of $W/W_f$. $a = 10$, $b = 8$.

## 2.2. *Evaporation rate*

To maintain a pinned contact line within the lubrication approximation the evaporative mass flux must be zero at the contact line for liquid drop. In our model we used a formula that was suggested in Ref. 15:

$$J(r,t) = J_0 \frac{1 - \exp\left(-A(1-r/R)^2\right)}{K + h(r,t)/h_0}. \tag{7}$$

We assumed the gel state must decrease the evaporation rate, $J$, so we inserted into the functional form of the evaporation rate (7) the factor with the functional form, for example, $f^{-1}$ (6).

## 3. Results

The model involves a set of nonlinear partial differential equations (4) with boundary and initial conditions (5), takes into account the equations (3, 6, 7) and requires numerical calculation for its analysis. Simulations were run for the dimensionless parameters: the capillary number is

$$Ca = \frac{\eta_0^2}{\varepsilon^3 \sigma \rho h_0} = 0.01;\ 0.1;\ 1,$$

the evaporation number

$$E = \frac{J_0 h_0}{\varepsilon \eta_0} = 0.1$$

and

$$\varepsilon = \frac{h_0}{R} = 0.06.$$

The initial volume fraction of the solute was $\frac{W_0}{W_f} = 0.1; 0.2; 0.3; 0.6$, $W_f = 0.5$, parameters: $a = 10$, $b = 8$, $A = 250$, $K = 1$. Our results are presented in the dimensionless form: the drop height was scaled by the $h_0$ and the radial coordinate $r$ was scaled by the $R$. The time evolution of the drop height profile is shown in Fig. 2 for: $t = 0$; $0.3t_f$; $0.5t_f$; $0.6t_f$; $0.7t_f$; $0.8t_f$; $t_f$; $1.5t_f$; $2t_f$. Here $t_f$ is the time when the final drop profile forms.

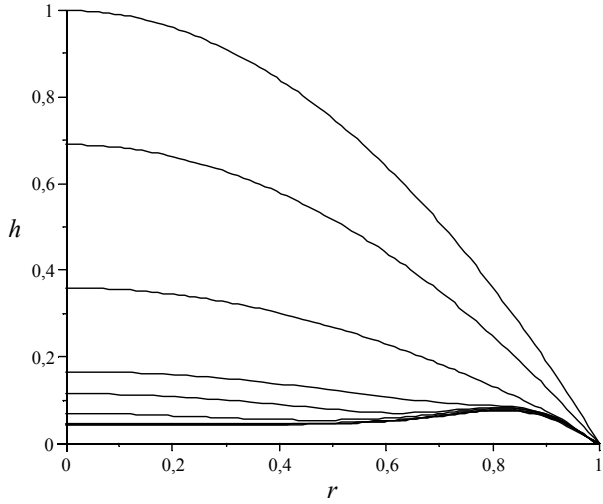

Figure 2. Time evolution of the drop height profile (from top to bottom) $W_0/W_f = 0.1$, $Ca = 0.1$.

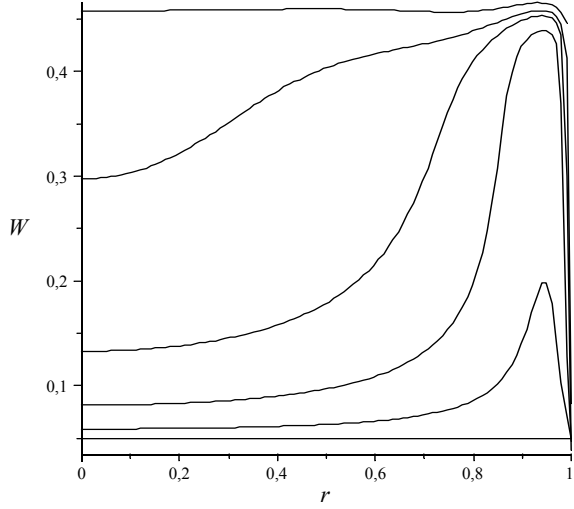

Figure 3. The height averaged volume fraction $W(r,t)$ as a function of $r$ (from bottom o top). $W_0/W_f = 0.1$, $Ca = 0.1$.

Numerical results for the time evolution of the height averaged volume fraction of the solute $W(r,t)$ as function of $r$ are shown in Fig. 3. The different curves correspond to $t = 0;\ 0.3t_f;\ 0.5t_f;\ 0.6t_f;\ 0.8t_f;\ t_f$. We can see the height averaged volume fraction of the solute $W$ increases with time. At early times the increase near the contact line is great and the value $W$ amounts to $0.9W_f$ at time $0.5t_f$ here. At later times, the value $W$ increases very slowly near the contact line and does not amount to value $W_f$ for the given parameters of the model. The drop height profile decreases and does not vary after the value $W$ reaches to about $0.8W_f$. Thus, we can see the deposit ring profile formation. The value $W \to W_0$ at $r \to R$ in our model as in the Ref. 15 that is the natural consequents of the utilized evaporative flux form (7).

The effect of the initial volume fraction $W_0$ of the solute on the drop height profile at time $t_f$ is shown in Fig. 4. For $W_0/W_f < 0.2$ the final drop height profile has deposit ring near the contact line. The width and height of the ring increase with the initial concentration of the solute. In our model, the flow is not capable of transferring all of the solute to the contact line, so the liquid zone of the drop contains some solute quantity at all time and there is gel zone in the central drop part for all investigated values $W_0$.

The variation of the capillary number $Ca$ ($Ca = 0.01$, $Ca = 1$) has not a visible effect on the results.

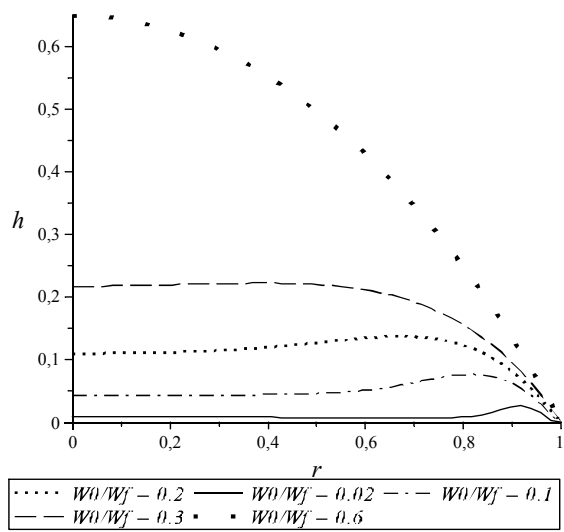

Figure 4. The drop height profile at time $t_f$ for the different initial volume fraction of the solute $W_0$. $Ca = 0.1$.

## 4. Conclusion

In this paper we investigated processes inside an evaporating sessile two-phase drop. We took into consideration that the viscosity of a colloidal solution is a function of volume fraction of the colloidal particles dispersed in the solution. Unfortunately, to our best knowledge the experimental data regarding surface tension and evaporating flux for the similar systems has not published yet. As a result, we are not in a position to perform a direct quantitative comparison with any experiments

## Acknowledgments

This work was supported by the Russian Foundation for Basic Research, project no. 09-08-97010-r povolzhje a.